\documentclass[final,1p,times]{elsarticle}

\usepackage{graphics}
\usepackage{color}
\usepackage{soul}

\usepackage{amssymb}
\usepackage{amstext}
\usepackage{array}

\biboptions{comma,square,sort&compress}
\newcommand{\isotope}[2]{$^{\rm #2}$#1}

\def	\evis	{E_{\rm vis}}
\def	\nhit	{N_{\rm hit}}

\def	\ave	{``AVE''}
\def	\min	{``MIN''}
\def	\max	{``MAX''}
\def	\Yexp	{Y_{\rm exp}}
\def	\Ypois	{Y_{\rm Pois}}
\def	\Yf	{Y_{\rm f}}
\def	\Cf	{C_{\rm f}}

\def	\ns	{{\rm ns}}
\def	\cm	{ {\rm cm}}

\def	\AMeV	{ {\rm MeV/nucleon}}
\def	\AGeV	{ {\rm GeV/nucleon}}

\def	\elab	{E_{\rm LAB}=600\ \AMeV}	
\def	\eff	{0.73}			 	

\journal{Nuclear Instruments and Methods in Physics Research A}

\begin{document}
\begin{frontmatter}

\title{Neutron recognition in the LAND detector\\ for large neutron multiplicity}

\author[IFJ]	{P.~Paw{\l}owski}
		\corref{corr}
		\cortext[corr]{Corresponding author. Tel.: +48 126628422, fax: +48 126628458}
		\ead{piotr.pawlowski@ifj.edu.pl}
\author[IFUJ]	{J.~Brzychczyk}
\author[GSI]	{Y.~Leifels}
\author[GSI]	{W.~Trautmann}
\author[IBJ,GSI]{P.~Adrich}
\author[GSI]	{T.~Aumann}
\author[IPN]	{C.~O.~Bacri}
\author[IFUJ]	{T.~Barczyk}
\author[ISF]	{R.~Bassini}
\author[GSI]	{S.~Bianchin}
\author[ISF]	{C.~Boiano}
\author[GSI]	{K.~Boretzky}
\author[CEA]	{A.~Boudard}
\author[GANIL]	{A.~Chbihi}
\author[IFJ]	{J.~Cibor}
\author[IFJ]	{B.~Czech}
\author[LNS]	{M.~De Napoli}
\author[CEA]	{J.-E.~Ducret}
\author[GSI]	{H.~Emling}
\author[GANIL]	{J.~D.~Frankland}
\author[CEA]	{T.~Gorbinet}
\author[GSI]	{M.~Hellstr\"om}
\author[LANL,GSI]{D.~Henzlova}
\author[SAS]	{S.~Hlavac}
\author[LNS]	{J.~Imm\`e}
\author[ISF]	{I.~Iori\footnotemark[2]}
\author[Goteborg,GSI]	{H.~Johansson}
\author[GSI]	{K.~Kezzar}
\author[IFUJ]	{S.~Kupny}
\author[CEA]	{A.~Lafriakh}
\author[GSI]	{A.~Le~F\`evre}
\author[CEA]	{E.~Le Gentil}
\author[CEA]	{S.~Leray}
\author[IFJ,GSI]{J.~{\L}ukasik}
\author[GSI]	{J.~L\"uhning}
\author[MSU]	{W.~G.~Lynch}
\author[GSI]	{U.~Lynen}
\author[IFUJ]	{Z.~Majka}
\author[MSU]	{M.~Mocko}
\author[GSI]	{W.~F.~J.~M\"uller}
\author[IBJ]	{A.~Mykulyak}
\author[GSI]	{H.~Orth}
\author[GSI]	{A.~N.~Otte}
\author[GSI]	{R.~Palit}
\author[CEA]	{S.~Panebianco}
\author[ISF]	{A.~Pullia}
\author[LNS]	{G.~Raciti\footnotemark[2]}
\author[LNS]	{E.~Rapisarda}
\author[GSI]	{D.~Rossi}
\author[CEA]	{M.-D.~Salsac}
\author[GSI]	{H.~Sann\footnotemark[2]}
\author[GSI]	{C.~Schwarz}
\author[GSI]	{H.~Simon}
\author[GSI]	{C.~Sfienti}
\author[GSI]	{K.~S\"ummerer}
\author[MSU]	{M.~B.~Tsang}
\author[MSU]	{G.~Verde}
\author[SAS]	{M.~Veselsky}
\author[CEA]	{C.~Volant}
\author[MSU]	{M.~Wallace}
\author[GSI]	{H.~Weick}
\author[GSI]	{J.~Wiechula}
\author[IFUJ]	{A.~Wieloch}
\author[IBJ]	{B.~Zwiegli\'{n}ski}
\footnotetext[2]{Deceased}

\address[IFJ]	{Institute of Nuclear Physics, PAN, Radzikowskiego 152, 31-342 Krak\'ow, Poland}
\address[IFUJ]	{Institute of Physics, Jagiellonian University, Reymonta 4, 30-059 Krak\'ow, Poland}
\address[GSI]	{GSI Helmholtzzentrum f\"ur Schwerionenforschung GmbH, D-64291 Darmstadt, Germany}
\address[IPN]	{Institut de Physique Nucl\'eaire, IN2P3-CNRS et Universit\'e, F-91406 Orsay, France}
\address[ISF]	{Istituto di Scienze Fisiche, Universit\'a degli Studi and INFN, I-20133 Milano, Italy}
\address[CEA]	{IRFU/SPhN, CEA/Saclay, F-91191 Gif-sur-Yvette, France}
\address[GANIL]	{GANIL, CEA et IN2P3-CNRS, F-14076 Caen, France}
\address[LNS]	{Dipartimento di Fisica e Astronomia - Universit\`a and INFN-CT and LNS, I-95123 Catania, Italy}
\address[MSU]	{Department of Physics and Astronomy and NSCL, MSU, East Lansing, Michigan 48824, USA}
\address[IBJ]	{National Centre for Nuclear Research, PL-00681 Warsaw, Poland}
\address[SAS]	{Institut of Physics, Slovak Academy of Sciences, Dubravska cesta 9, 84511 Bratislava 45, Slovakia}
\address[LANL]	{Los Alamos National Laboratory, New Mexico, USA}
\address[Goteborg] {Fundamental Fysik, Chalmers Tekniska H\"ogskola, 412 96 G\"oteborg, Sweden}

\begin{abstract}
The performance of the LAND neutron detector is studied. Using an event-mixing
technique based on one-neutron data obtained in the S107 experiment at the GSI
laboratory, we test the efficiency of various analytic tools used to determine
the multiplicity and kinematic properties of detected neutrons. A new algorithm
developed recently for recognizing neutron showers from spectator decays in the
ALADIN experiment S254 is described in detail. Its performance is assessed in
comparison with other methods. The properties of the observed neutron events are
used to estimate the detection efficiency of LAND in this experiment.
\end{abstract}


\end{frontmatter}

\section{Introduction}
The Large Area Neutron Detector (LAND) was constructed and used at
the GSI Laboratory in Darmstadt for the measurement of neutrons from
reactions at relativistic energies \cite{LAND}. It consists of 200 modules
(paddles) of 2~m length and 10x10~cm$^2$ cross section, constructed from
interleaved 5~mm thick iron and plastic-scintillator plates. They are arranged
in 10 consecutive planes of 20 paddles each, alternating between horizontal and
vertical paddle orientation, to form a compact 2x2x1~m$^3$ detector.

The interaction of a neutron with the detector material results in the
production of light pulses (hits) in one or more paddles. Photomultipliers
register the light at both paddle ends, providing information on the amplitude
and time of occurrence of each hit.
The hit position along the paddle is determined from the time difference of the
two signals. Each hit generated by a neutron is thus characterized by its
position in space and time.
Typical internal resolutions of the position and time measurements are approximately
$\pm 3~\cm$, and $\pm 0.25~\ns$, respectively \cite{yordanov}.

The measured amplitudes are summed into the
so-called ”visible energy”, i.e. the energy deposited in the scintillator
material of the detector. It is only a small part
of the total energy deposited by an interacting neutron, as the majority of it is
lost in the iron layers, invisible for the phototubes. Charged particles
impinging on the detector are recognized by their signals in a VETO wall of 5-mm
thick plastic-scintillator slabs mounted in front of the detector \cite{LAND}.

LAND was mainly intended to serve as a time-of-flight spectrometer with the precise
determination of the position and arrival time of identified neutrons
being the primary goal. Because of its calorimetric properties, it has also been used
to study multi-neutron emissions in specific experiments.
For determining the multiplicity of detected neutrons, two calorimetric
observables can be used: the hit multiplicity $\nhit$ and the total visible
energy $\evis$, provided the average values of these quantities per incident
neutron are known \cite{Zude}. To obtain also the kinematic properties of
individual neutrons, a more complicated data treatment is needed. It requires
the identification of the first hit generated by the neutron, called here
”primary hit”. Its registered location and time can be used to determine the
velocity vector of the incident neutron. For neutron (or, more generally,
particle) multiplicities exceeding one, the individual hit patterns may overlap,
turning the identification of primary hits into a complex problem. In fact, if
the overlapping hit patterns are also generated close in time, unique solutions
may not always be possible. A procedure for an optimized primary-hit recognition
on an event-by-event basis is thus a necessary tool for experiments requiring
multi-neutron detection, but also an interesting challenge. The assessment of its
performance under the given experimental conditions will be essential.

The standard neutron recognition algorithm used so far \cite{Keller} seems to be successful in
experiments where the neutron multiplicity to be expected is low. They include
structure studies of halo nuclei as, e.g., \isotope{Li}{11} by fragmentation at 280~{\AMeV}
\cite{Zinser} or the excitation of collective modes as, e.g.,
the two-phonon giant dipole resonance in heavy nuclei \cite{Boretzky} or the
pygmy dipole resonance in neutron-rich nuclei \cite{Klimkiewicz}. The same
algorithm was also used for analyzing the data from the collective-flow
experiments providing evidence for the neutron squeeze-out in
\isotope{Au}{197} + \isotope{Au}{197} collisions at 400~{\AMeV} \cite{Leifels, Lambrecht}
even though the most probable neutron multiplicity was of the order of 3 to 4,
with nearly the same number of light charged particles being present in the recorded events.
Because of considerably larger neutron multiplicities, in the \isotope{U}{238} + Fe at 1~{\AGeV}
experiment, the neutron yields were measured with the use of only two forward LAND planes,
with a very limited neutron detection efficiency \cite{yordanov}.

Since similarly large neutron multiplicities were expected in
the more recent study of projectile fragmentation at 600~{\AMeV}, performed
with beams of \isotope{Sn}{107,124}, \isotope{La}{124} on \isotope{Sn}{nat} targets
\cite{Pawlowski, Trautmann11},
a new algorithm for neutron-shower recognition was developed. It exploits the hit-to-hit time
and space correlations expected for a shower and is intended to be especially
adapted to the high neutron multiplicities from the de-excitation of highly
excited projectile residues. The algorithm has been already used for the LAND data analysis
in the S248@GSI and S304@GSI experiments performed by the SPALADIN Collaboration
\cite{LeGentil, Gorbinet}. It will also be available for the
data collected in the very recent study of the density dependence of the
symmetry energy by comparing the collective flows of neutrons and light charged
particles from heavy-ion collisions at 400~MeV/nucleon. In this experiment of
the ASYEOS Collaboration, the LAND detector served as the main tool for the
measurement of the emissions of neutrons and hydrogen isotopes near mid-rapidity
\cite{Russotto, ASYEOS}.

Considering the wide range of applications and the complexity of the problem, it
is indispensable to acquire a quantitative knowledge of the performance of the
used algorithm under the conditions of the particular experiment. Since
ambiguous hit patterns without unique solutions have to be expected, this is
even more important and of general interest for users of LAND. The present work
is intended to serve this purpose by performing a comparative test of the
standard and new algorithms with synthetic events generated from single neutron
data from deuteron breakup. This experimental data set had been collected 20 years ago
by measuring the response of LAND with tagged neutrons generated by dissociating
deuterons of energies 70 to 1050~{\AMeV} and by recording the protons
deflected by the ALADIN magnet (experiment S107@GSI, see ref. \cite{S107, Boretzky}).
Earlier tests with the same technique of generating synthetic multi-neutron
events had shown that the quality of neutron detection and kinematic
identification changes with the experimental conditions \cite{Cub}. It is clear
that it will also depend on the actual status of readiness of the detector;
missing channels, e.g., will deteriorate the performance. Therefore, to make the
simulation as realistic as possible, the measured hit
distribution in time-space obtained in the \isotope{Sn}{124} + \isotope{Sn}{nat}
reaction at $\elab$, and the state of readiness of LAND in the frame of the S254@GSI experiment,
\cite{Trautmann08,Sfienti,Trautmann09,Ogul} are taken into account.

Even though this work is especially devoted to the LAND detector, the methods developed here
  can also be considered in the wider context of large neutron detectors. They should be useful
  for the NeuLAND detector proposed within the frame of the R3B project at the future FAIR facility
\cite{R3B}

\section{\label{Procedures}Neutron recognition procedures}
\subsection{\label{Standard}Shower Volume Algorithm (SVA)}
The standard algorithm for neutron recognition in LAND, Shower Volume Algorithm, has
been already described
in \cite{Keller, Cub, Zinser}. Here we present a short description of the method, to
give the reader an idea about
the differences between the two algorithms studied in the paper.

In SVA the main focus is put on the time sequence of registered hits. The first
registered hit is a primary hit, defining properties of the first detected
neutron. Subsequently registered hits are attributed to the same neutron if they
are located in a cylindrical volume of 29 ~cm radius and 50~cm depth around the
primary hit, or at a larger distance, if they fulfill the kinematic conditions
of quasi-free scattering. The hits attributed to the neutron in this way
are removed from further analysis. Then the same treatment is applied to the
remaining hits. This will select a number of primary hits which, in the ideal
case, will correspond to the neutrons that have hit the detector.

The cylindrical volume used in this procedure is quite large. Although it
amounts to only 3~\% of the total LAND volume, one should remember that in a
real experiment the neutron angular distribution may be strongly anisotropic
so that only a small part of LAND is irradiated by the majority of neutrons. For
this reason, the SVA algorithm starts to fail when the neutron multiplicity
increases: the cylindrical volume frequently contains hits from more than only one neutron
shower. As a consequence, the resulting neutron multiplicity becomes
underestimated.
Moreover, as the hit-time order plays an important role,
the faster neutrons are favored by the algorithm. In the events, where the neutron
multiplicity is considerably large, this fact may disturb the observed neutron velocity
distribution, underestimating the yields in the range of lower velocities.
These disadvantages of the algorithm were the reasons for searching
for a more refined algorithm devoted especially to the reactions characterized by
a large neutron multiplicity.

\subsection{\label{New}Shower Tracking Algorithm (STA)}
As mentioned above, a high-energy neutron or a charged particle, interacting with the
LAND detector, produces usually several hits. To correctly determine the kinematic
properties of the incoming particle one has to recognize properly the first
("primary") hit. Its time and position determine the velocity of the particle.
The consecutive secondary hits are correlated in time and space with the primary
hit, or a previously identified secondary hit, forming a kind of "jet" or
"chain" of hits. Such a secondary hit identification can be done using recursive
programming methods. This is the main idea of the Shower Tracking Algorithm.

\begin{figure}[p]
\begin{centering}
\includegraphics[width=8cm]{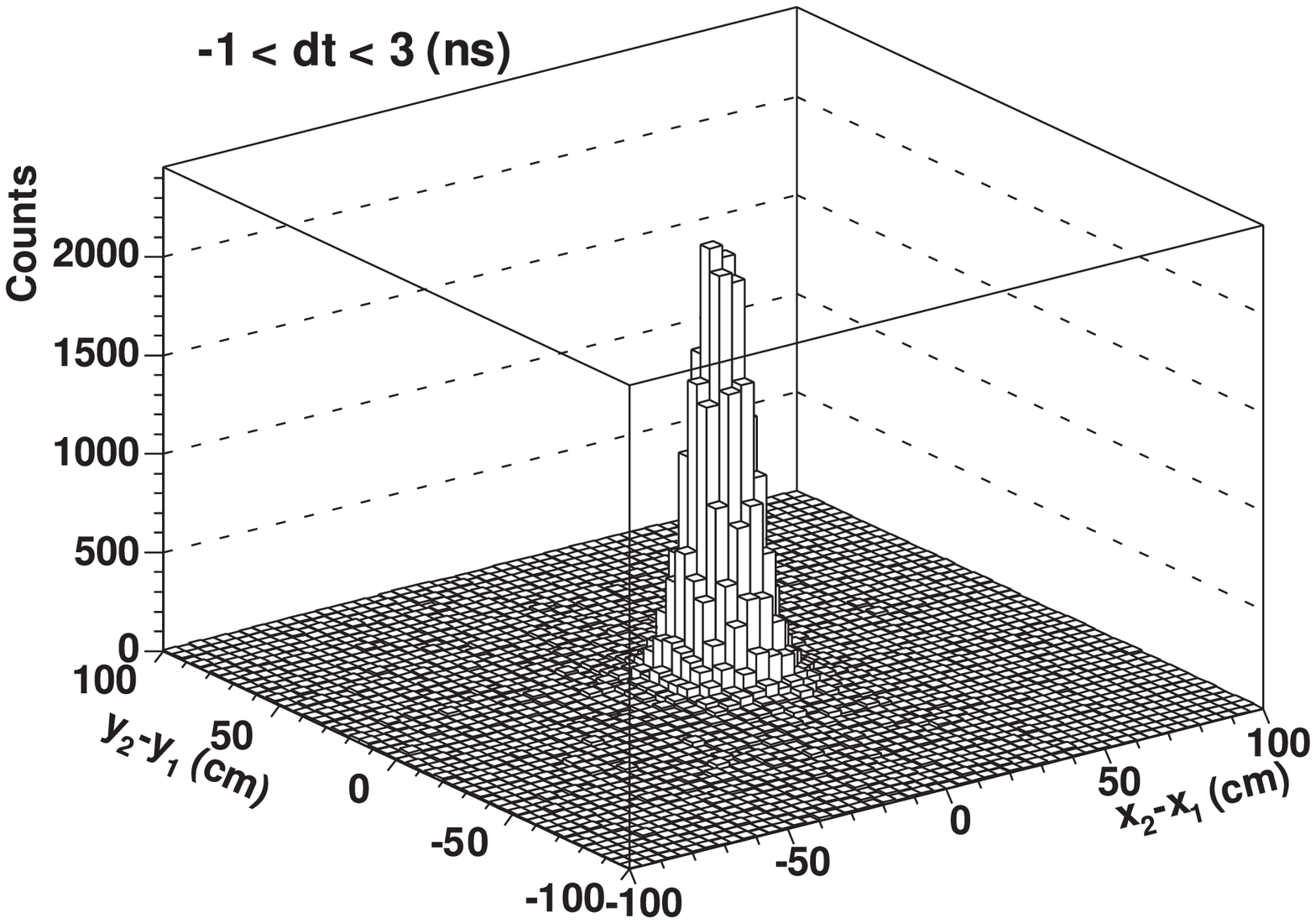}
\par\end{centering}
\caption{\label{correlation}A typical hit-hit correlation for two hits
registered in two neighboring planes and in the time interval
$-1\,\ns<t_{2}-t_{1}<3\,\ns$
(experimental data from S254@GSI)}
\end{figure}

Let us consider two hits observed in two neighboring LAND planes, with time-space
coordinates $(x_{1}, y_{1}, t_{1})$ and $(x_{2}, y_{2}, t_{2})$.
If the two hits are produced by the same interacting particle, they should be
relatively close to each other. Indeed, this is clearly observed when one studies plots of
$y_{2}-y_{1}$ vs. $x_{2}-x_{1}$  (Fig.~\ref{correlation}).

Analyzing hit-hit correlations one can define the correlation condition:
\begin{equation}
\begin{array}{c}
-19\,\cm<\Delta x<19\,\cm,\\
-19\,\cm<\Delta y<19\,\cm,\\
-1\,\ns<\Delta t<3\,\ns.\end{array}
\label{main_window}
\end{equation}
Using this condition one can identify a ``cluster'' of hits generated by a
particle. A recursive procedure starts from a hit registered in the first LAND
plane.
For this hit, a search is made for a correlated hit in the same and in the next plane.
If a correlated hit is found, a search for a hit correlated with this secondary hit is made.
The procedure is continued until no more correlated hits are found. All correlated hits
are marked as secondary hits and removed from the total pool of hits. Then the procedure
starts from the beginning with one of the remaining hits,
descending from the first plane to the second plane,
third plane, and so on.

After the clustering procedure has been completed, a number of ``single'' hits remain, where no
correlation
with other hits was found. Some of them are really ``singlets'',
due to a small momentum transfer
between the reacting neutron and nuclei, or when the reaction  occurs close
to the edge of the LAND detector.
But a number of them is a result of detection problems. In this number one
should mention:
\begin{enumerate}[(i)]
\item \label{paddles}Missing paddles - a number of LAND paddles are not
working (or improperly working) in the experiment.
In this case some hit chains are broken.
\item \label{multi-hits}Multi-hits - when two hits are registered in one paddle,
they are seen as one hit with a false position.
Also its time is earlier than the real times.
\item \label{longrange}The neutrons can interact with LAND in two points,
distant from each other. These two points may be seen by the procedure as two separated
clusters but should be attributed to one neutron.
\end{enumerate}
We are not able to take properly into account the points (\ref{paddles}) and
(\ref{multi-hits}). We can only estimate their influence
on the neutron multiplicity. To do this, we construct two limiting
procedures. The first one, called {\max}, maximizes
the neutron multiplicity: it treats all the remaining single hits as separate clusters.
The second one, called {\min}, minimizes the multiplicity
of neutrons, trying to clusterize the remaining single hits using a looser
condition - only a wide time correlation is here taken
into account:
\begin{equation}
-7\,\ns<\Delta t<4\,\ns.\label{time_only}
\end{equation}

A third procedure, called {\ave}, is proposed to get a probably more realistic value
in between these extremes. Here, for the remaining single hits, a condition is used
that is more restrictive than (\ref{time_only}) but still less restrictive
than (\ref{main_window}):
\begin{equation}
\begin{array}{c}
-19\,\cm<\left(\Delta x \ {\rm or }\ \Delta y\right)<19\,\cm, \\
-1\,\ns<\Delta t<3\,\ns,\end{array}
\label{time_and_longitude}
\end{equation}
i.e. only the $x$ or the $y$ separation is required to meet the correlation criterion.
In the {\min} and {\ave} procedures, the conditions (\ref{time_only}) and (\ref{time_and_longitude}),
respectively, are applied in place of (\ref{main_window}) in the main clustering procedure, and a
second run of the procedure is performed for the ensemble of single hits.

To take into account the point (\ref{longrange}) we use a ``long-range
correlation procedure'' searching for distant clusters correlated
in time and space. For each previously determined primary hit we use its measured time and position to estimate the velocity of the incoming neutron.
Then we search for another primary hit in a cylinder defined by this
velocity vector as the cylinder axis,
and a fixed radius of 19 cm. Such a hit is treated as correlated with the first
hit, if its time fulfills the relation:
\begin{equation}
\left|\Delta t-\frac{d}{v}\right| \le\ 1 \ns,
\label{long-range-time}
\end{equation}
where $d$ is the distance between the two primary hits and $v$ is the velocity of the
first primary hit. This treatment is added as the last step
in the procedures {\ave} and {\min}.

In the following, the STA acronym refers to the  {\ave} procedure, which will be shown
to be the most accurate. The {\min} and {\max} procedures are considered
as error estimators.

\section{\label{Tests}Testing efficiency of neutron recognition}
\subsection{\label{Mixing}Mixing single-neutron data}
The validation of the neutron recognition procedures has been performed by
analyzing synthetic multi-neutron events generated
from single-neutron data provided by the S107@GSI experiment \cite{S107}. In the
experiment the deuteron beam was
scattered on a \isotope{Pb}{208} target at energies from 70 to 1050 {\AMeV}. The
experimental set-up was focused on deuteron break-up registration.
The neutrons hitting LAND were measured in coincidence with the accompanying
proton, deflected by the magnetic field of the ALADIN magnet. In this way
the information about the showers produced by a single neutron in LAND  was
collected.
From this experiment, an efficiency curve for neutron detection with LAND was obtained
\cite{Boretzky, Cub}. For neutrons of 600 MeV, e.g., a detection efficiency of 94\%
was observed with the fully operational detector. Moreover, the single-neutron
events registered in LAND can be randomly mixed at random positions, simulating in
this way two-, three-, or more-neutron events with predetermined time and spatial
distributions.

We used this technique to determine the accuracy of the neutron recognition
algorithms and the systematic and statistical errors generated
by various analytic tools. In the mixing procedure we take into account the
treatment used before in \cite{Zude} for multiple hits in one paddle:
\begin{enumerate}
\item A new time is calculated by taking into account only the signals arriving first at
the two photo-tubes, and then the new hit time and position are derived (this is a slight
approximation to the actual response of the electronic readout when the signals overlap);
\item The signal energies of the two hits are summed at both paddle ends, and
then the visible energy is recalculated.
\end{enumerate}

Furthermore, the actual spatial and temporal hit distributions, as measured in
S254@GSI experiment, are reproduced in the mixed
data. Also non-working paddles in this experiment have been removed from simulated
events.

\subsection{\label{Dynamics} Resolving efficiency and quality}
In the evaluation of a recognition algorithm one should consider two main
problems: some neutron showers detected in LAND
are lost due to low time-space resolution of the applied algorithm - i.e. two or
more showers are erroneously recognized as a single one. On the other hand, in some cases
the algorithm produces "false" showers, i.e. a one-neutron shower is artificially identified
as two or more showers - because of either inadequacies of the algorithm, or
some detection problems. From the point of view of neutron recognition, only
the assignment of the primary hit is important. A mistake in attributing secondary
hits to appropriate showers does not play any role.

Besides the correct determination of the actual multiplicities, the recognition of the
individual neutron showers is also important. A minimum requirement is that all hits,
identified as primary, are actually belonging to different showers.
In the simulated multiple neutron data it is possible to check
the origin of each primary hit recognized by an algorithm.
If a number $n$ of them belong in fact to the same neutron shower (i.e. the same
S107@GSI event) we speak about one "true" neutron and $n-1$ "false" neutrons.
If a neutron shower does not have a representative among the primary hits,
we call this neutron "lost".
\begin{figure}[ht]
\begin{centering}
\includegraphics[width=8cm]{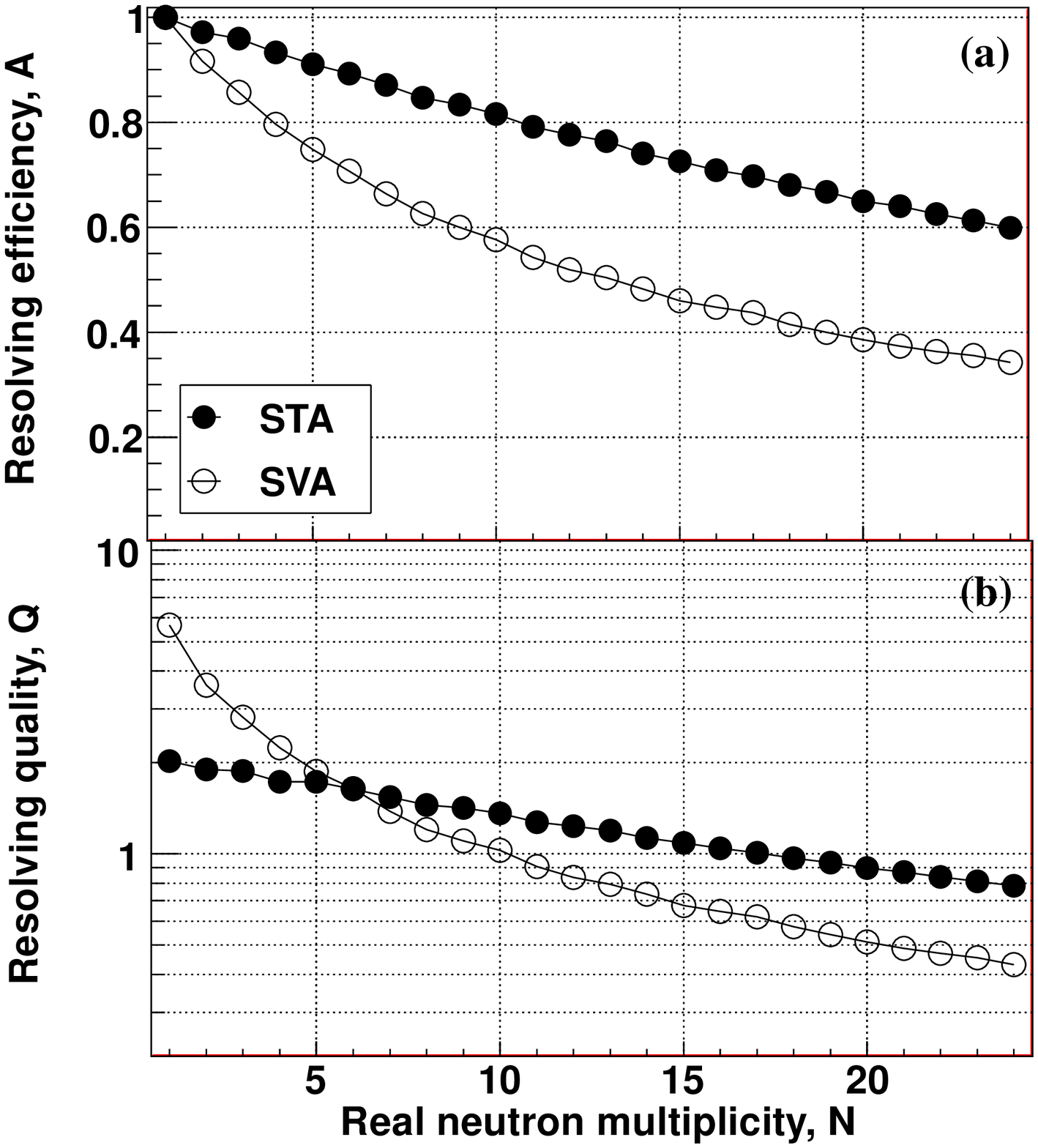}
\par\end{centering}
\caption{\label{resolving} Resolving efficiency (a) and quality (b) obtained for
STA (closed points) and SVA (open points) algorithms, as functions of the real neutron
multiplicity. The missing neutron fraction (see Section \ref{eff}) is not taken into account.
The simulation uses actual spatial and temporal hit distributions
measured at $\elab$ incident energy in the S254@GSI experiment}
\end{figure}

Let us denote the number of true neutrons as $N_T$, and the number of lost
neutrons as $N_L$. Then the total number of initial neutrons $N$
can be expressed by the sum:
\begin{equation}
N_{T} + N_{L} = N.
\end{equation}
On the other hand, the total number of recognized primary hits $N_{C}$ can be
divided into two groups: the number of true neutrons $N_T$, and the number of
false neutrons $N_F$:
\begin{equation}
N_{T} + N_{F} = N_{C}.
\end{equation}
Let us define the variables:
\begin{eqnarray}
A &=& \frac{ N_{T}}{N}, \\
Q &=& \frac{N_{T}}{N_{F}+N_{L}}.
\end{eqnarray}
The resolving efficiency $A$ expresses the rate of correctly identified neutrons.
It is evident from Fig.~\ref{resolving}a that, independent of $N$, STA is more efficient
in recognizing true showers than SVA. But this is not a sufficient argument to favor STA.
Let us imagine an algorithm treating each registered hit as a separated neutron shower.
Its resolving efficiency will be unity for all $N$, because $A$
does not carry the information about the number of false showers
$N_{F}$. A more useful variable for characterizing an algorithm is the resolving quality $Q$.
The larger its value is, the better the algorithm works. Fig.
\ref{resolving}b shows the dependence of this quantity on neutron multiplicity.
At low neutron multiplicities, the resolving quality of SVA is higher but it decreases
dramatically when the multiplicity becomes larger. Above the crossing point at $N=6$
the resolving quality of STA is superior. For this reason, STA should be applied
in experiments with high neutron multiplicity.
When the expected neutron multiplicity does not exceed
6, SVA appears to be the better choice.

\subsection{\label{Multiplicity} Neutron multiplicity estimation}
We consider here four estimators for the  determination of the neutron multiplicity
on an event-by-event basis:
\begin{enumerate}[(a)]
  \item\label{nhit} Normalized hit multiplicity $\nhit/\mu$,
  \item\label{evis} Normalized total visible energy $\evis/\epsilon$,
  \item\label{sva} Primary hit multiplicity given by SVA,
  \item\label{sta} Primary hit multiplicity given by STA.
\end{enumerate}
In observables (\ref{nhit}) and (\ref{evis}), $\mu$ and $\epsilon$ are,
respectively, the mean hit multiplicity and the mean visible
energy generated by one incoming neutron. These parameters can be
determined from one-neutron data. Unfortunately, the results of the
S107@GSI experiment cannot be directly used in other experiments, as specific
setup conditions (in particular CFD thresholds and geometry) can considerably
change these mean values.
Alternatively, one can apply SVA and/or STA algorithms to select one-neutron
events from the studied experimental data.
We apply this latter concept to determine the values $\mu$ and $\epsilon$
and we adopt the average of the two values given by the SVA and STA algorithms.
Such a treatment used with the S107@GSI data has given the results which differed
only by few percent from the actual $\mu$ and $\epsilon$ derived directly from the data.

\begin{figure}[ht]
\begin{centering}
\includegraphics[width=7cm]{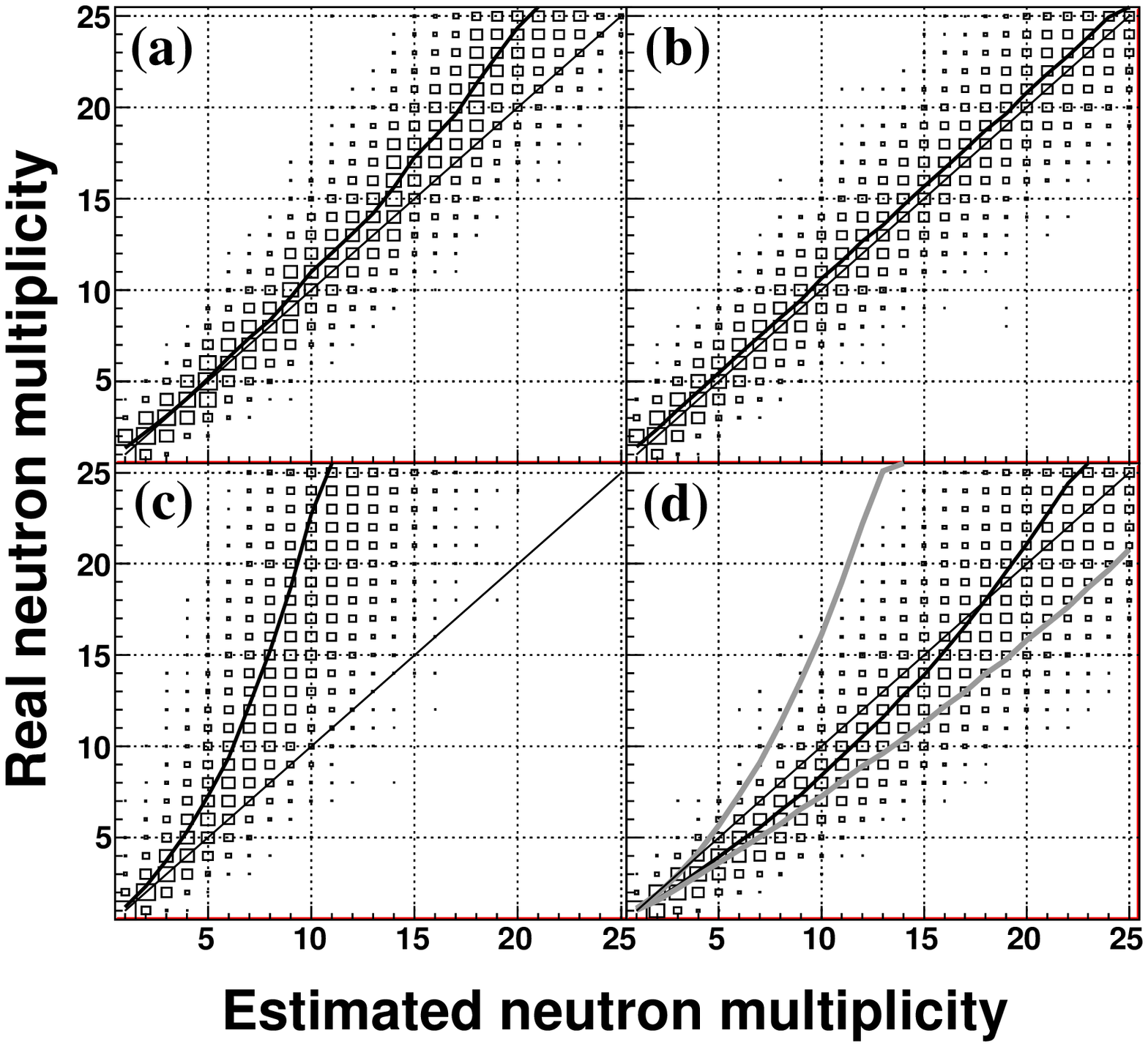}
\par\end{centering}
\caption{\label{estimators}Comparison of neutron multiplicity estimators: (a)
$\nhit/\mu$, (b) $\evis/\epsilon$,
(c) SVA, (d) STA. The solid lines show the average neutron multiplicity as a
function of the estimated value.
The boxes represent the event distributions. The gray lines in (d) correspond to
the {\min} and {\max} limiting procedures. The thin solid line shows the 1:1
correspondence case. The simulation uses actual spatial and temporal
hit distributions measured at $\elab$ incident energy in the S254@GSI experiment}
\end{figure}

The estimated neutron multiplicities are compared with the known multiplicities of
synthetic multi-neutron events in Fig.~\ref{estimators}. The solid lines represent
the average real multiplicity as a function of the estimated value. The distribution
of the symbols reflects the uncertainties of the procedures. The evidently most accurate
result is obtained with the estimator based on the $\evis$ observable.
This is a simple consequence of the fact that in the mixing of one-neutron events we simulate
the multi-hit visible energy by summing all partial energies. If this model is
correct, the total visible energy is retained in the case of double hits,
whereas the number of hits does not grow if a paddle sees two hits. The resulting
upbend of the true vs. estimated multiplicity can be calibrated but the resolution
becomes poor at high multiplicity (Fig.~\ref{estimators}a). It is also evident
that the proportionality factor $\epsilon$ is quite well
determined by the recognition algorithms.

In the lower row of Fig.~\ref{estimators} one can compare the performance of the
two recognition algorithms. It is evident that SVA
fails in neutron multiplicity determination when the multiplicity is larger than 4.
\begin{figure}[ht]
\begin{centering}
\includegraphics[width=7cm]{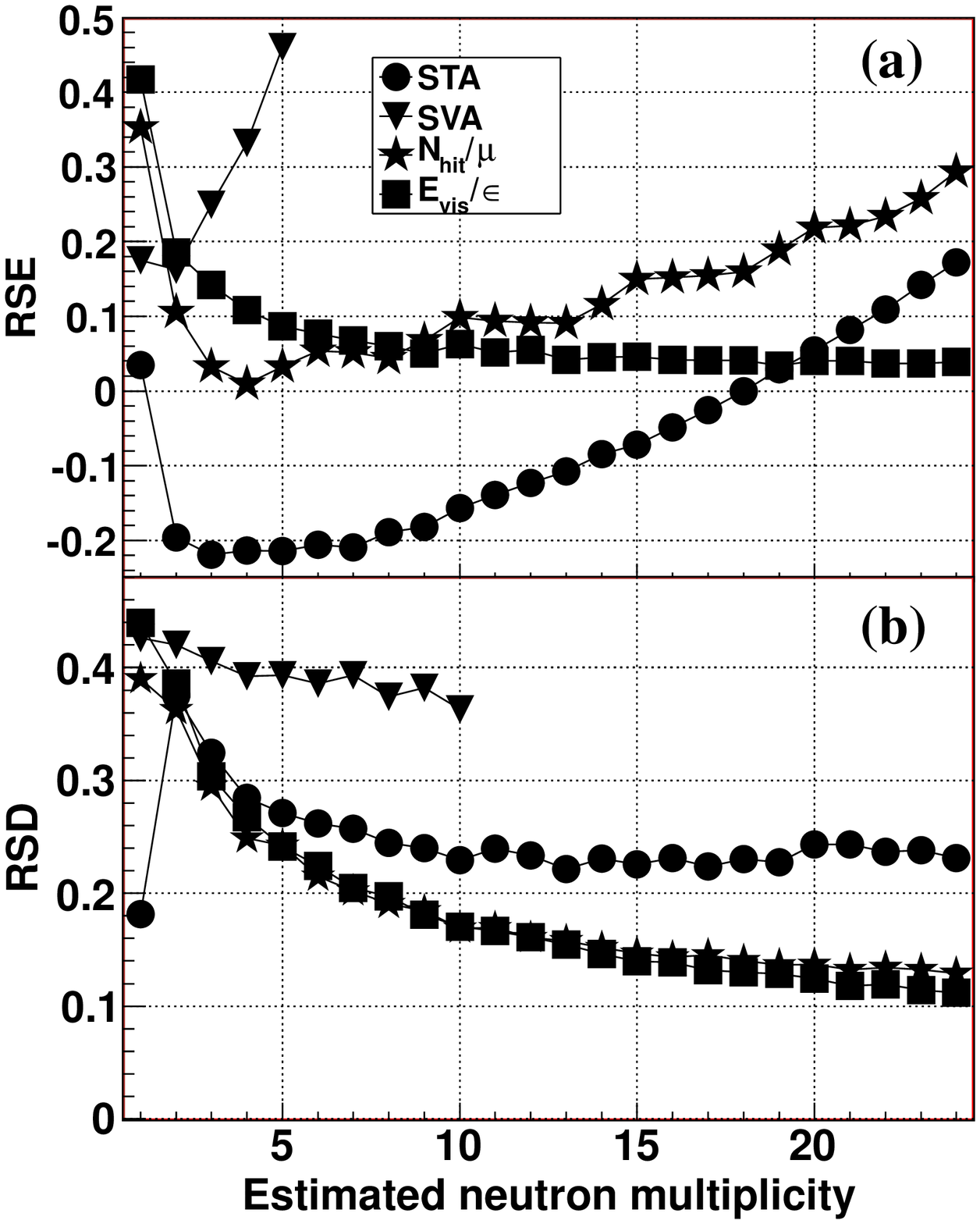}
\par\end{centering}
\caption{\label{errors}Comparison of relative systematic errors \textit{RSE} (a)
and relative statistical deviations \textit{RSD} (b)
generated by neutron multiplicity estimators, as functions of the estimated neutron multiplicity.
The simulation uses actual spatial and temporal hit distributions measured at $\elab$
incident energy in the S254@GSI experiment}
\end{figure}

A more precise study of the errors generated by the estimators is presented in
Fig.~\ref{errors}.
The relative systematic error (\textit{RSE}) and relative standard deviation (\textit{RSD})
can be expressed as follows:
\begin{eqnarray}
\textit{RSE} & = & \frac{\overline{N}-N_{c}}{N_{c}},\\
\textit{RSD} & = &
\frac{\sqrt{\overline{N^{2}}-\overline{N}^{2}}}{\overline{N}}.
\end{eqnarray}
Fig.~\ref{errors}a shows the comparison of  \textit{RSE} as a function of the estimated
neutron multiplicity. It can be seen that the calorimetric estimators give
relatively better accuracy for larger multiplicities.
A similar result can be observed in Fig.~\ref{errors}b where \textit{RSD} is presented.
In the range of lower multiplicities, the recognition algorithms give better
accuracy in comparison with the calorimetric estimators.
For larger multiplicities $\evis$ gives the best estimate.

Concluding, when only the total neutron multiplicity is needed, the $\evis/\epsilon$
estimator gives the best measure.
A typical systematic error in an event-by-event analysis is lower than $\pm$10~\%.
It corresponds to an absolute
neutron multiplicity deviation better than $\pm1$. The statistical errors are below 25 \%
for multiplicity 8 and larger. To improve the predictions in the range of
 small multiplicities one can combine this estimator with the others.

\begin{figure}[ht]
\begin{centering}
\includegraphics[width=7cm]{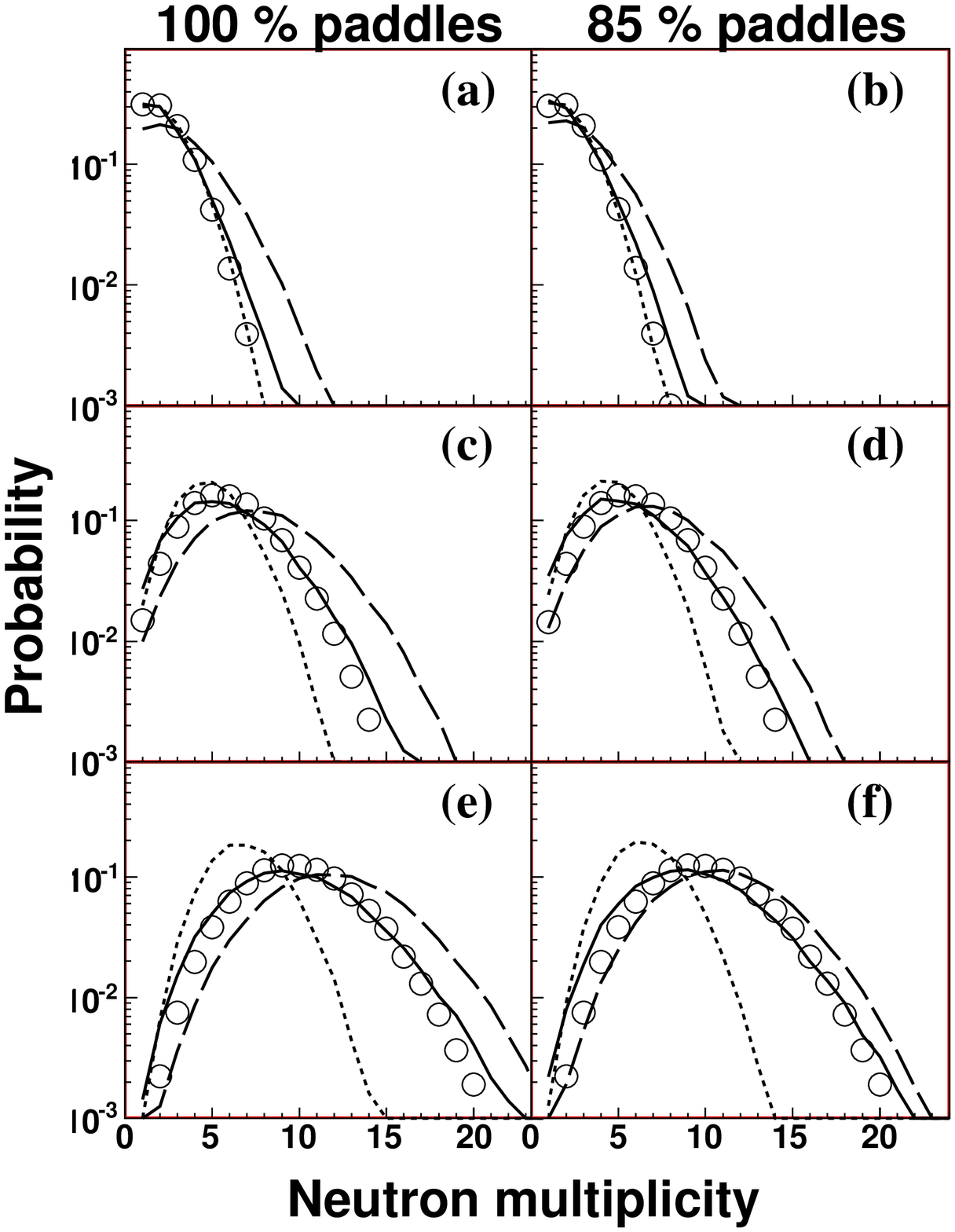}
\par\end{centering}
\caption{\label{distro}Simulated Poissonian neutron distributions (circles) and
reconstructed distributions
using $\evis/\epsilon$ estimator (solid line), STA (dashed line) and SVA (dotted
line). The left column corresponds
to full LAND performance, the right column to only 85\% working paddles. The mean
neutron multiplicities are:
$\overline{N}=2$ (upper row). $\overline{N}=6$ (middle), and  $\overline{N}=10$
(bottom). The simulation uses actual spatial and temporal hit distributions
measured at $\elab$ incident energy in the S254@GSI experiment}
\end{figure}

Let us study the predictive power of the estimators for an entire multiplicity
distribution. Using mixed neutron data one can simulate
a Poissonian multiplicity distribution and compare it to the distribution
reconstructed with different multiplicity estimators.
Fig.~\ref{distro} shows the results of such a simulation done for the neutrons
from the S254@GSI experiment at 600 ~{\AMeV}. Evidently, SVA reproduces
the multiplicity distribution very well if the mean multiplicity is around 2. When
the multiplicity increases, this algorithm is less efficient,
and the use of STA will be more appropriate. The best estimate is again provided by $\evis/\epsilon$.

In an actual experiment, a part of the LAND paddles may not be working. This was the
situation in the S254@GSI experiment, where about 15~\%
of LAND paddles were out of order, mainly in the rear part of the detector.
The right column of Fig.~\ref{distro} shows that the obtained multiplicity distributions
are very similar, even in the case of such a defect.

\section{\label{Efficiency}Neutron detection efficiency}
The efficiency of LAND for neutron detection is high for high neutron energies.
For neutrons from the deuteron breakup at 600~{\AMeV}, a value $\eta_{0} = 0.94$
was measured in the S107@GSI experiment with the then fully functional detector.
In a given experiment, the actual detection efficiency may deviate from this value
because of a specific geometry, the aging of the scintillator material\footnote{The reduction
of the average attenuation length from originally 215~cm \cite{LAND} to 116~cm at the time of
the S254 experiment, is evidence for a degradation of the detector material with time.}
and possibly modified thresholds, or non-working paddles. The efficiency is also expected
to change near the edges of the detector. In the following subsections, these aspects will be
addressed.

\subsection{\label{High-efficiency-area}High efficiency area}

\begin{figure}[ht]
\begin{centering}
\includegraphics[width=8cm]{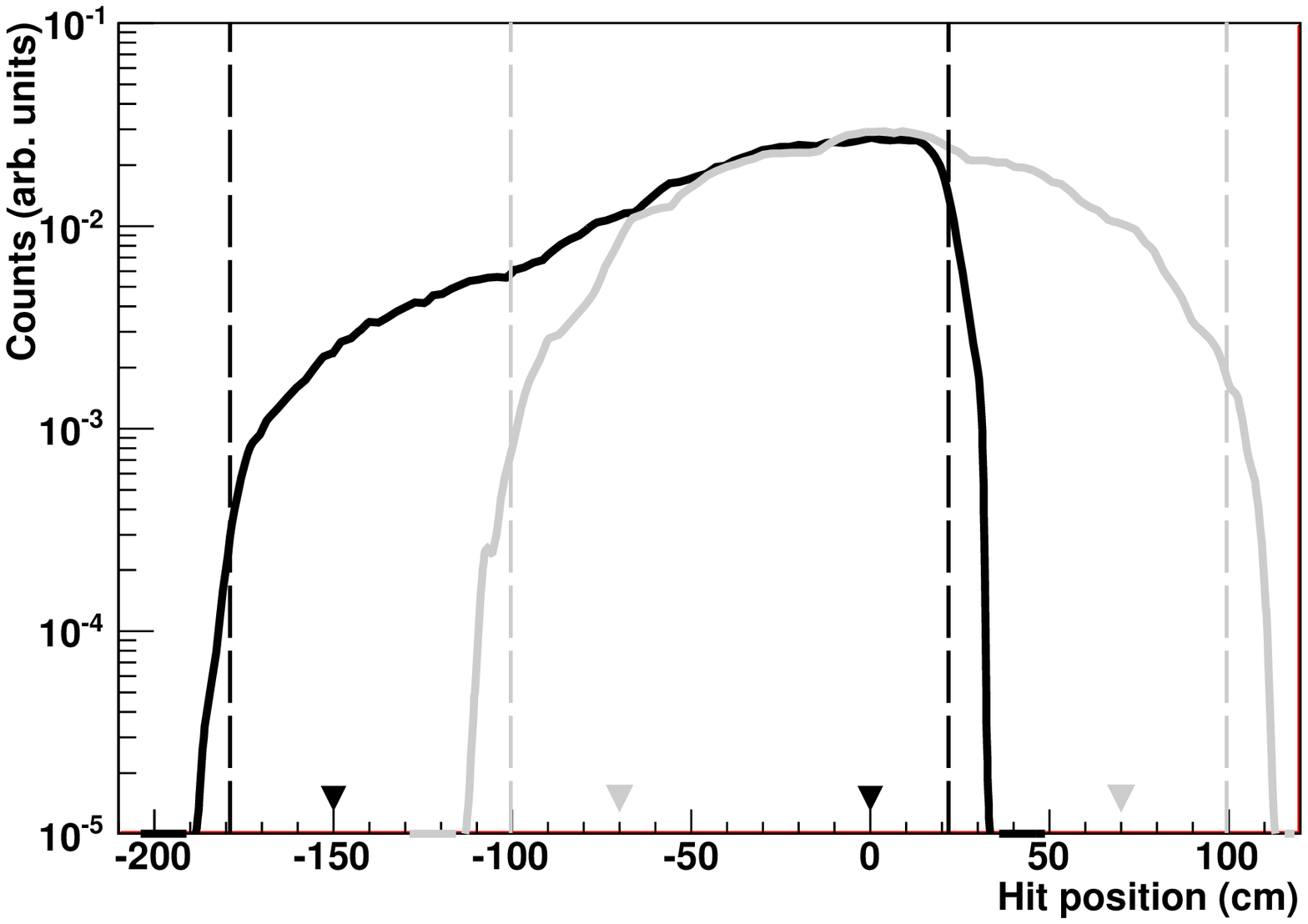}
\par\end{centering}
\caption{\label{high-efficiency}Distribution of absolute hit positions along a paddle for
horizontal (black line) and vertical (gray line) paddles,
as observed in the S254@GSI experiment. Dashed vertical lines show the corresponding geometrical limits
of the detector.
Triangles indicate the software limits (\ref{high_eff_area}) imposed on the data to get a
homogeneous efficiency area. The recorded hit-position distributions exceed the actual
dimensions of LAND because of the finite resolution of the time measurement}
\end{figure}

The S254@GSI experiment had the layout of instrumentation nearly identical to
that presented in Fig.~1 of Ref. \cite{Schuttauf}. The LAND detector was positioned
approximately 10 m downstream from the target, vertically symmetric with respect to
the beam axis but displaced horizontally by 78~cm.
It resulted in a symmetric vertical ($y$ direction) but strongly
asymmetric horizontal ($x$ direction) distribution of hit positions.
The magnetic field of ALADIN does not affect the neutron trajectories,
and the neutron emission is expected to have a cylindrical symmetry,
with similar hit distributions in both vertical and horizontal directions.
Because of the finite acceptance of the ALADIN magnet \cite{Schuttauf},
shadowing effects near the edges of  LAND are expected.
Examples of the horizontal and vertical hit distributions obtained from the horizontally
and vertically oriented paddles, respectively, are shown in Fig.~\ref{high-efficiency}
in the beam-centered coordinate frame. The expected azimuthal symmetry is observed
up to a distance of about 70~cm from the center. The more rapid vertical decrease
beyond 80~cm coincides with the shadow of the ALADIN poles and magnet chamber as projected
from the target position. The horizontal distribution extends further out.
Here the projected shadow of the magnet is close to the end of the detector.
Both distributions start to drop more rapidly at about 15~cm from the end of the detector
but extend beyond it because of the finite position resolution obtained from the time
differences at the paddle ends.

Also the yield observed on the opposite end of the horizontal paddles (at $+20~\cm$)
is lower than that observed in the corresponding section of vertical ones.
It suggests some additional edge effect, maybe connected to lower detection efficiency
at the ends of a paddle.

Because of these obvious edge effects and in order to work with a homogeneous detection
efficiency we select a high efficiency area by cutting off the edges of the detector.
This area is defined by:
\begin{equation}
\begin{array}{c}
-150\,\cm<x<0\,\cm,\\
-70\,\cm<y<70\,\cm.
\end{array}
\label{high_eff_area}
\end{equation}

\subsection{\label{eff}Estimation of missed neutron fraction}

\begin{figure}[ht]
\begin{centering}
\includegraphics[width=7cm]{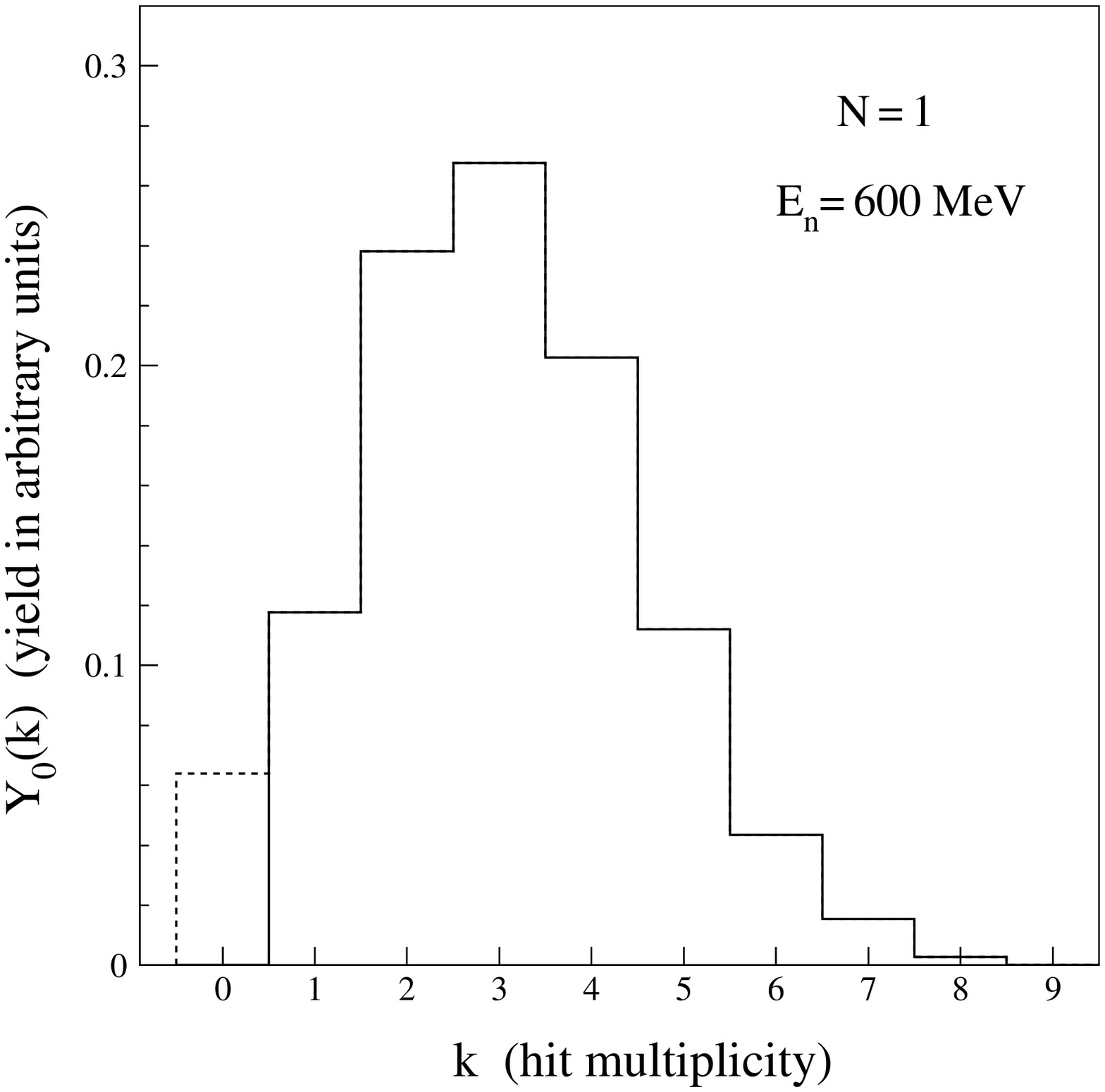}
\par\end{centering}
\caption{\label{ideal_detector}The hit multiplicity distribution for the fully operational
detector (solid line) and estimated fraction of missed neutrons (dashed line).
Result obtained from S107@GSI experiment}
\end{figure}

The distribution of the number of hits $Y_{0}(k)$, generated by single neutrons of energy
600~MeV from the S107@GSI data and for the fully operational detector, is shown in Fig.
\ref{ideal_detector} by the solid histogram. The measured efficiency for this energy
is $\eta_{0}=0.94$ \cite{Boretzky}, indicating that 6\% of the incident neutrons pass
the detector without a visible interaction. The corresponding probability $Y_{0}(0)=0.06$
is given by the dashed line in Fig.~\ref{ideal_detector}.

\begin{figure}[ht]
\begin{centering}
\includegraphics[width=7cm]{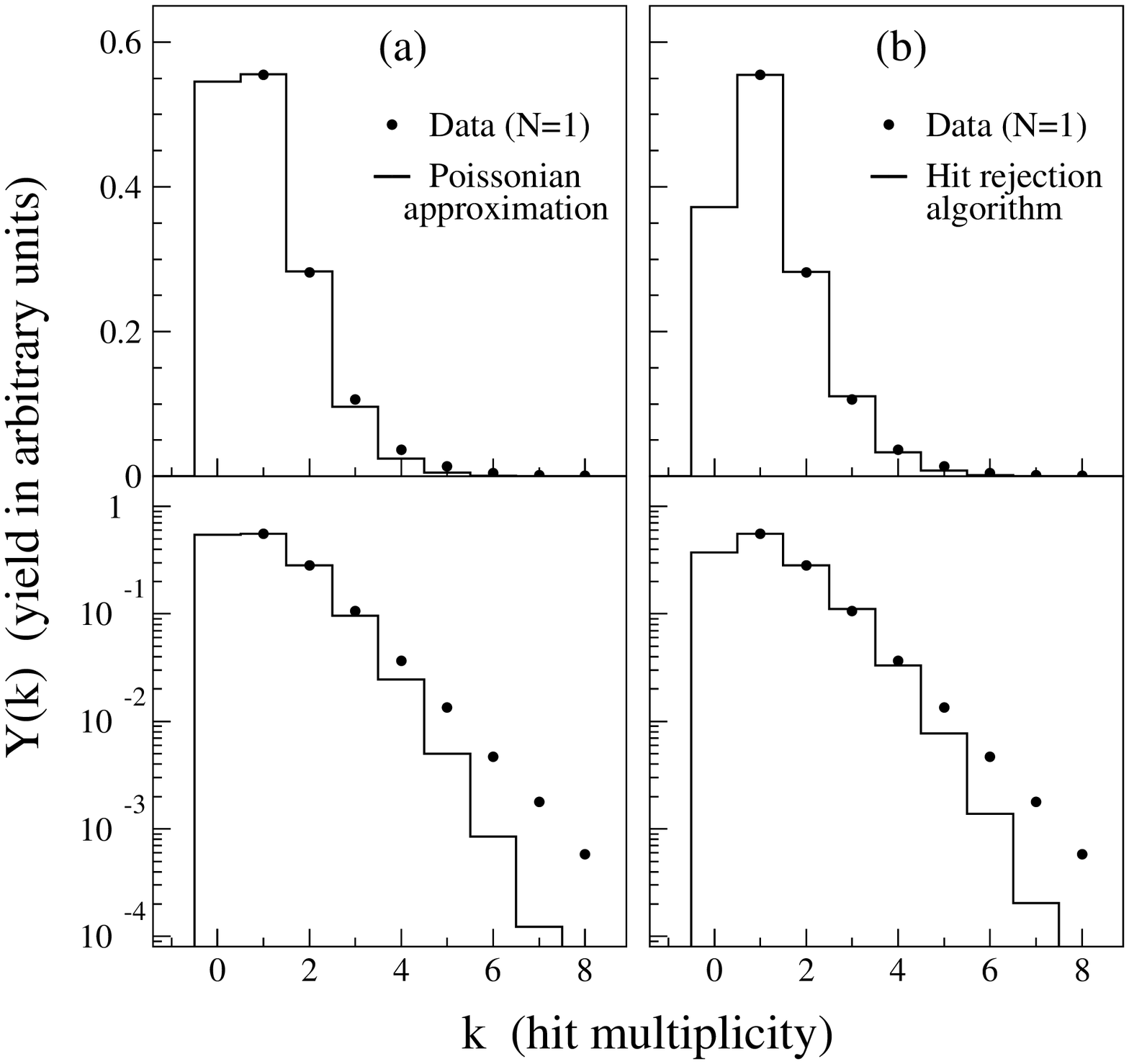}
\par\end{centering}
\caption{\label{real_detector}The experimental hit multiplicity distribution
for neutron multiplicity $N=1$ with its approximations by a Poissonian distribution (a)
and a distribution from the hit rejection procedure (b). The lower
panels show the same in log scale}
\end{figure}

The hit multiplicity distribution for $N=1$ in the S254@GSI experiment,
$\Yexp(k)$, obtained by selection of single neutron events with the procedure
{\ave} (with primary hit located in the \emph{high efficiency area}) is represented
by the points in Fig.~\ref{real_detector}.
Comparing with $Y_{0}(k)$, the distribution is much narrower, the average number
of hits per neutron is almost halved. This significant reduction
of the hit multiplicity is mainly due to non-working segments of the
detector $(\sim15\%)$ and a considerably higher hit registration
threshold.

The estimation of the detector efficiency requires an extrapolation of the hit
distributions to include their value for $k=0$, assuming a realistic functional form
of the distribution.

First, let us apply the Poissonian distribution
$\Ypois(k)=C\frac{\lambda^{k}}{k!}e^{-\lambda}$
with two adjustable parameters: the normalization constant $C$ and
the mean value $\lambda$. The parameters are determined from the
conditions $\Ypois(k)=\Yexp(k)$ for $k=1$ and $k=2$, that provide
the best approximation in the low multiplicity range. The resulting
$\Ypois(k)$ distribution with $\lambda=1.02$ is shown in
Fig.~\ref{real_detector}a by the histogram.
With the assumption $\Yexp(0)=\Ypois(0)$ the procedure suggests the
efficiency $\eta = 1.0-\exp(-1.02)=0.64$.

Alternatively, one may try to determine $Y(k)$ with a model closer to the actual
detector situation. We use the distribution $Y_{0}(k)$ measured with the fully operational
detector as a primary distribution which is then modified by rejecting a certain
fraction of hits.

The hits are considered as a linear chain. Each hit in the chain is accepted or rejected with
a given rejection probability $P$. If the chain is broken by rejected hit(s) into two or
more separate parts, such an event is removed from the considered sample of events
with $N=1$ (the clustering procedure would
recognize two or more neutrons). After applying this random rejection
method the final distribution is found as:
\begin{eqnarray}
\Yf(k=0) & = & \Cf\sum_{i\geq{0}}P^{i}Y_{0}(i), \label{Yf}\\
\Yf(k\geq1) & = &
\Cf(1-P)^{k}\sum_{i\geq{k}}(i+1-k)P^{i-k}Y_{0}(i).
\end{eqnarray}
Requiring $\Yf(k)=\Yexp(k)$ for $k=1$ and $k=2$ gives $\Cf$ and
$P=0.507$ with the result shown in Fig.~\ref{real_detector}b.
Using the expression for $k = 0$ (Eq. (\ref{Yf})),
the efficiency is found to be $\eta=\eff$ for this particular experiment.
This is significantly lower than the value of 94~\% previously deduced from the
deuteron-breakup experiment S107@GSI performed with the then new and fully functional detector.
The estimated detection efficiency is specific for the S254@GSI experiment and may be different
in other experiments. With the presented method, it can be obtained from the measured data.

\section{Conclusions}
An overview has been presented on the analytic tools, applied so far in the analysis of
the LAND data. We compare two calorimetric observables and two different neutron recognition
algorithms. The work estimates the systematic and statistical errors for an event-by-event
neutron multiplicity estimation. The new algorithm, recently developed for the S254@GSI experiment,
is described in detail and compared to the algorithm used before. A discussion of the applicability
of each algorithm is presented. The new algorithm seems to be a good choice for the analysis of the
data, when expecting neutron multiplicities larger than 4, especially when the emission of both
neutrons and protons has to be studied.

The study allowed to estimate the neutron-detection efficiency of the LAND to be about 70~\%.
It is worth noting that this estimate is significantly lower
than the value 94~\% deduced many years ago from the results of the deuteron-breakup
experiment S107@GSI. The difference is believed to be due to the natural aging of the
scintillator material, possibly higher effective thresholds of the discriminators,
and to the about 15~\% non-working paddles. This emphasizes the need for methods for
determining the detection efficiency from the actual data measured in a specific experiment.
Significant variations can, apparently, not be excluded and will be difficult to model
since some of the required parameters may not be sufficiently well known. The methods presented
here can be expected to serve this purpose.

\section*{Acknowledgments}
This work has been supported by the Polish Ministry of Science and Higher
Education under Contracts No. PB 1 P03B 020 30 (2006-2009), N202 160 32/4308 (2007-2010),
DPN/N108/GSI/2009, and the European Community under Contract No. HPRI-CT-1999-00001.

\bibliographystyle{elsarticle-num}
\bibliography{references}

\end{document}